\documentclass[aps,pre,twocolumn,showpacs,superscriptaddress,groupedaddress]{revtex4}

\usepackage{graphicx}  % needed for figures
\usepackage{hyperref}
\usepackage{bm}% bold math
\usepackage{color}

\newcommand\Rm{{\rm Rm}}
\newcommand\Pm{{\rm Pm}}
\newcommand\Ru{{\rm Re}}

\begin{document}

%\begin{frontmatter}

\title{MHD Turbulence in spin-down flows of liquid metals}


\author{Peter~Frick}
\affiliation{Institute of Continuous Media Mechanics,  Korolyov str.1, Perm, 614013, Russia}
\affiliation{Perm State University, Bukireva str. 15, Perm, Russia}
\author{Irina Mizeva }
\affiliation{Institute of Continuous Media Mechanics,  Korolyov str.1, Perm, 614013, Russia}

%\author{Peter~Frick$^{1,2}$}
%\author{Irina~Mizeva$^{1}$\corref{mycorrespondingauthor}}
%\ead{mizeva@icmm.ru}
%\address{$^2$ Institute of Continuous Media Mechanics,  Korolyov 1,Perm, 614013, Russia}
%\address{$^2$ Perm State University, Bukireva str. 15, Perm, Russia}

\begin{abstract}

Intense spin-down flows allow one to reach high magnetic Reynolds numbers in relatively small laboratory setups using moderate mass of liquid metals. The spin-down flow in toroidal channels was the first flow configuration used for studying dynamo effects in {\it non-stationary} flows. In this paper, we estimate the effect of small-scale dynamo in liquid metal spin-down flows realized in laboratory experiments (Denisov S., Noskov V., Stepanov P. and Frick P., JETP Letters \textbf{88}, (2008); Frick P., Noskov V., Denisov S. and Stepanov R., PRL \textbf{105}, (2010); Noskov V., Denisov S., Stepanov. R and Frick P., PRE \textbf{85}, (2012)). Our simulations have confirmed the conclusion that the dynamo effects observed in the experiments done on gallium are weak -- a slight burst of small-scale magnetic energy arises only at the highest available rotation velocity of the channel. In sodium flows, the induction effects are quite strong -- an essential part of kinetic energy of sodium spin-down flows is converted into magnetic energy and dissipates because of Joule heat losses. We have extended our simulations beyond the capabilities of existing laboratory facilities and examined the spin-down flows at the channel rotation velocity $\Omega \gg 50$rps. It has been found that $\Omega\approx 100$rps is enough to reach the equipartition of magnetic and kinetic spectral power density at the lowest wave numbers (largest scales), whereas at $\Omega \gtrsim 200$rps the intensity of the magnetic field becomes comparable to the intensity of velocity field fluctuations. We have also studied the influence of the magnetic Prandtl number on the efficiency of small-scale dynamo in spin-down flows. In the experimental spin-down flows, the small-scale dynamo remains in a quasi-kinematic regime, and magnetic energy is mainly dissipated at the same scale, wherein it is converted from kinetic energy. The real small-scale dynamo  starts to operate at $\Pm>10^{-4}$, and the inertial range of the magnetic energy spectrum appears. Thereupon the energy dissipation is postponed to a later time and smaller scales, and the peak of turbulent energy (both kinetic and magnetic) slightly increases with $\Pm$.
\end{abstract}


%\begin{keyword}
%turbulence, shell model
%\end{keyword}

%\end{frontmatter}
\maketitle

\section{Introduction}
\label{sec:intro}

Induction effects become observable in flows of conductive fluids if the characteristic magnetic Reynolds number achieves the order of unity. Actually, the magnetic Reynolds number $\Rm=UL/\eta$  ($U$ is the characteristic velocity, $L$ is the characteristic size, and $\eta$ is the magnetic diffusivity) defines the ratio of the generation term to the dissipation one in the equation for magnetic field induction; thus $\Rm \approx 1$ means that the induction effects are of order of the dissipation effects.
The ordinary (hydrodynamic) Reynolds number $\Ru =  UL/\nu$ ($\nu$ is the kinematic viscosity) is related to $\Rm$ through the magnetic Prandtl number $\Pm  =\nu/\eta = \Rm / \Ru$, which  is very low for liquid metals ($10^{-5}$ and less).
 Therefore, a real challenge for dynamo related experiments is the generation of laboratory liquid metal flows characterized by even moderate (of order 1 or more) magnetic Reynolds numbers \cite{Stefani2008,2014PhyU...57..292S}. The maintenance of flows capable to support a stationary MHD-dynamo requires huge laboratory facilities \cite{Gailitis2000,2001PhFl...13..561S,Monchaux2007}.

Giving up the requirement of stationarity, one can realize a dynamo-suitable regime in a strong pulsed flow of moderate mass of liquid metal. The intense spin-down screw flow, generated by abrupt braking of a fast rotating toroidal channel with special diverters inside, has been proposed as a laboratory model for dynamo experiments \cite{Denisov1999,Frick2002}. The MHD experiments, in which the spin-down flow of liquid gallium was investigated, allowed one to observe the alpha-effect produced by a joint action of the large-scale vorticity and the gradient of turbulent pulsations  \cite{Stepanov2006}.
The study of the spin-down flow dynamics in a torus without diverters has revealed that the development of the flow in the channel is attended by a strong short-time burst of turbulent pulsations \cite{Noskov2009}. This burst of small-scale turbulence gives chance to observe an increase in the effective resistivity of liquid metal. Such experiments were realized in the spin-down flows of gallium \cite{Denisov2008} and sodium \cite{Frick2010PRL} and made it posiible to estimate effective magnetic diffusivity in turbulent flows.

The evolution of non-stationary turbulent flows in the braked tori was studied in detail using local dual-axis potential probes for the toroidal and poloidal velocity components in a thin torus filled with gallium (radius of the cross-section $r = 0.0225$m, median radius of the torus $R= 0.0875$ m, $r/R = 0.257$, $\Ru < 6 \cdot 10^5$, $\Rm < 1$) \cite{Noskov2009} and in a thick torus filled with sodium ($r = 0.08$m, $R= 0.18$ m, $r/R = 0.44$, $\Ru < 2 \cdot 10^6$, $\Rm < 20$)  \cite{Noskov2012}. However, direct measurements of the magnetic field inside the metal flow were inaccessible, and the studies of the magnetic field were limited by some integral induction effects, such as the effective conductivity of turbulent metal flow \cite{Denisov2008,Frick2010PRL} or turbulent diamagnetism \cite{Frick2015MH}.

Planning non-stationary experiments in the toroidal flow, which reaches the dynamo threshold in a very short time (about a second), researchers supposed that this time would be enough to obtain the large-scale screw dynamo effect \cite{Frick2002}. However, the laboratory screw dynamo has to operate in a highly turbulent flow, where the small-scale dynamo should also arise. The small-scale dynamo  has been widely studied in stationary forced MHD turbulence \cite{2005ApJ...625L.115S,2010PhRvE..82e6326U}, as well as in free decaying turbulence \cite{Frick2010,Brandenburg2015PRL}. How far can  the small-scale magnetic field develop in a very short pulse flow? Will this time be sufficient to reach the equipartition at least in some part of the spectrum? These questions were not addressed neither theoretically nor experimentally. Direct numerical simulation (DNS) of the nonstationary flow at $\Ru \approx 10^7$ is on the verge of capability of modern computer clusters. This made attractive the use of a reduced model of turbulent flow.

The aim of this paper is twofold. First, we make an attempt to apply a shell model of turbulence that is commonly used to model fully developed homogeneous MHD turbulence \cite{Plunian2013} for simulation of   small-scale turbulence in a real, very specific flow. Second, we investigate the dynamics and spectral structure of a turbulent small-scale magnetic field generated in real laboratory environment and some hypothetical turbulent spin-down flows.

\section{Spin-down flow}
\label{sec:mod}
We consider a spin-down flow in a fast rotating toroidal channel after its abrupt braking. $R$ is the median radius of the torus, and $r$ is the radius of its cross-section. The channel is filled by the liquid metal, which rotates like a solid body before braking and then is pushed forward by inertia during braking. The simplest model predicting turbulent flow evolution in this toroidal channel, introduced in \cite{Denisov1999}, relates the mean velocity of inertial motion along the channel, $V_{tor}$  to the tangential wall stress $\tau$
\begin{equation} {{d V_{tor}}\over{dt}} = -R{{d \Omega}\over{dt}}-{{2
\tau} \over {r \rho}}\,, \label{ac} \end{equation}
where the velocity is measured in the frame of reference associated with the channel, $\Omega$ is the angular velocity of the torus, $\rho$ is the fluid density, $\tau = \rho v_*^2$, and $v_*$ is the so-called dynamic velocity. For the smooth pipe and  $\Ru > 10^5$, the dynamic velocity is related to the mean velocity by an empirical law \cite{schlichting1979}
\begin{equation}
V_{tor}=v_*\left(2.5 \ln {{r v_*}\over\nu} +1.75\right).
 \label{Vv} \end{equation}
Eqs.~(\ref{ac}) and (\ref{Vv}) allows one to describe time evolution of the mean flow in the braked torus and the decay  of the spin-down flow after braking \cite{Frick2002}.

Below, we simulate the evolution of small-scale turbulence in real flows using $V_{tor}$ obtained  experimentally \cite{Noskov2009,Noskov2012} and use relation (\ref{Vv}) to calculate the dynamic velocity $v_*$.

\section{Shell model}
\label{sec:shell}

We use the shell model for MHD turbulence introduced in \cite{Mizeva2009} but rewritten in dimensional form
\begin{eqnarray}
d_t U_n={W_n}({\bf U},{\bf U})-(\rho\mu\mu_0)^{-1}{W_n}({\bf B},{\bf B})- \nu k_n^2 U_n + F_n, \nonumber \\
d_t B_n={W_n}({\bf U},{\bf B})-{W_n}({\bf B},{\bf U})-\eta k_n^2 B_n + B_E k_n U_n, \label{eq}
\end{eqnarray}
where  ${\bf U}=(U_0,...U_{N-1})$ and ${\bf B}=(B_0,...B_{N-1})$ are vectors in the space $C^N$, and $N$  is the total number of shells. The shell width is defined by $\lambda$, wave numbers  $k_n=k_0\lambda^n$, and the lowest wave number corresponds to the channel radius $k_0=2\pi/r$. The structure of equations (\ref{eq}) mimics the original MHD equations -- the bilinear form ${\bf W}({\bf X},{\bf Y})$ is like ${\bf X}\cdot\nabla{\bf Y}$, and the term $F_n$ describes the action of external forces in the shell $n$.  We use the $W_n({\bf X},{\bf Y})$ suggested
in \cite{Mizeva2009}, which can be rewritten in a general form as
\begin{eqnarray}
W_n({\bf X},{\bf Y})=ik_n[(X_{n-1}Y_{n-1}+X_{n-1}^*Y_{n-1}^*)-\lambda
X_n^*Y_{n+1}^*\nonumber\\
-\frac{\lambda^2}{2}(X_n Y_{n+1}+X_{n+1}Y_n+X_nY_{n+1}^*+X^*_{n+1}Y_n)\nonumber\\
-\frac{\lambda}{2}(X_{n-1}^*Y_{n-1}-X_{n-1}Y_{n-1}^*) +\lambda X^*Y_{n+1}] \nonumber
\\ -i k_n \lambda^{-5/2}[\frac{1}{2}(X_{n-1}Y_n+X_nY_{n-1})+\lambda
X^*_nY_{n-1}^*\nonumber\\
-\lambda^2(X_{n+1}Y_{n+1}+X_{n+1}^*Y_{n+1}^*)+\frac{1}{2}(X_nY_{n-1}^*+X_{n-1}^*Y_{n})
\nonumber\\
 -\lambda X^*_nY_{n-1}+\frac{\lambda}{2}(X_{n+1}^*Y_{n+1}-X_{n+1}Y_{n+1}^*)].\phantom{XX}
\label{W}\end{eqnarray}In the non-dissipative limit, the non-linear form (\ref{W}) provides the conservation of the total energy $E=\sum(E^u_n+E^b_n)$, cross-helicity $H^c=\sum(U_nB_n^*+B_nU_n^*)/2$ and magnetic helicity $H^b= \sum \imath k_n^{-1}((B_n^*)^2-B_n^2)/2$.

Turbulence is supposed to be generated by some mechanical force described in shell equations by the term $F_n$. We suggest that turbulence in the bulk flow is maintained by the wall stress acting at the largest scale ($n=0$). With an empirical parameter $\kappa$, it takes the form
 \begin{equation}
 F_0 = \kappa {2 v_*^2  \over  r}
 \label{kappa}
 \end{equation}
Note that shell models generally succeed in reproducing in detail many relative properties of fully developed turbulence, such as  energy distribution among the scales, the ratio of different mechanisms of spectral transfer or dissipation of energy, enstrophy, helicities, etc. However, these models fail to give absolute values and, hence, require some calibration. In our case, $\kappa$ is the adjustable parameter, which will be used for such calibration.

The last term in the induction equation describes  magnetic field generation due to the background (Earth's) magnetic field, which is $B_E\approx 0.5$Gs.

\section{Results}
\label{sec:res}

\subsection{Modeling laboratory experiments}

Using the available data of the laboratory experiments on spin-down flows of liquid gallium \cite{Noskov2009} and sodium \cite{Noskov2012}, we have calibrated our model and reproduced the evolution of velocity fluctuations in both spin-down flows. Some parameters of both experiments are given in Table~1.  Note that the shown Reynolds numbers $\Ru$ and $\Rm$ are estimated using the maximal values of the toroidal velocity. In the gallium experiment, this $\Rm$  was below unity, and the experiment corresponded to the very beginning of development of small-scale induction effects, which manifest themselves in a weak increase (about 1 \%)  of the measured effective magnetic diffusivity \cite{Denisov2008}. In the sodium flow, $\Rm$ reached 20, and even the magnetic Reynolds number, defined through the velocity fluctuations, exceeded unity  \cite{Noskov2012}.Then, the induction effects become stronger, and  the effective magnetic diffusivity increased by a factor of $\approx 1.5$ \cite{Frick2010PRL}.

\begin{table}[h]
\caption{Laboratory experiments} \label{tab:av1}
\begin{tabular}{|l|c|c|}
\hline
        & Gallium flow\cite{Noskov2009}   & Sodium flow \cite{Noskov2012} \\ \hline
 $\rho, kg/m^3$       &  $ 6095 $ & $ 928 $ \\
 $\nu,  m^2/s$       & $ 3\cdot10^{-7} $ &  $ 9\cdot10^{-7} $\\
 $\eta, m^2/s$     & $ 0.21 $ & $ 0.078 $  \\
 R, m        & $ 0.0875 $ & $ 0.18 $  \\
 r, m     & $ 0.0225 $ & $ 0.08 $  \\
 $\Omega_{max}$,  rps  & $55$ & $50 $\\
 $\Ru_{max}$ & $ 7\cdot 10^5 $ & $2\cdot 10^6 $\\
 $\Rm_{max}$ & $ 0.3 $ & $ 20 $ \\ \hline
\end{tabular}
\end{table}

The toroidal pipe flow is absolutely unstable and generates the poloidal mode at any Reynolds number \cite{Chupin2008}. The higher is the channel curvature (the ratio $r/R$), the stronger is the poloidal flow. In the gallium experiment  ($r/R = 0.257$) the poloidal velocity reached about 10\% of the toroidal velocity, while in the sodium experiment ($r/R = 0.44$) it reached about 18\%. It is worth noting that during the braking the burst of turbulence is characterized by strong dominance of poloidal pulsations. In both experiments at the end of braking the anisotropy of velocity fluctuations (defined as the ratio of rms of poloidal fluctuations to rms of toroidal fluctuations) reached about 2.5. The shell model (\ref{eq}) is introduced for isotropic turbulence and ignores the effect of anisotropy.

\begin{figure}
\begin{center}
\includegraphics[width=0.49\textwidth]{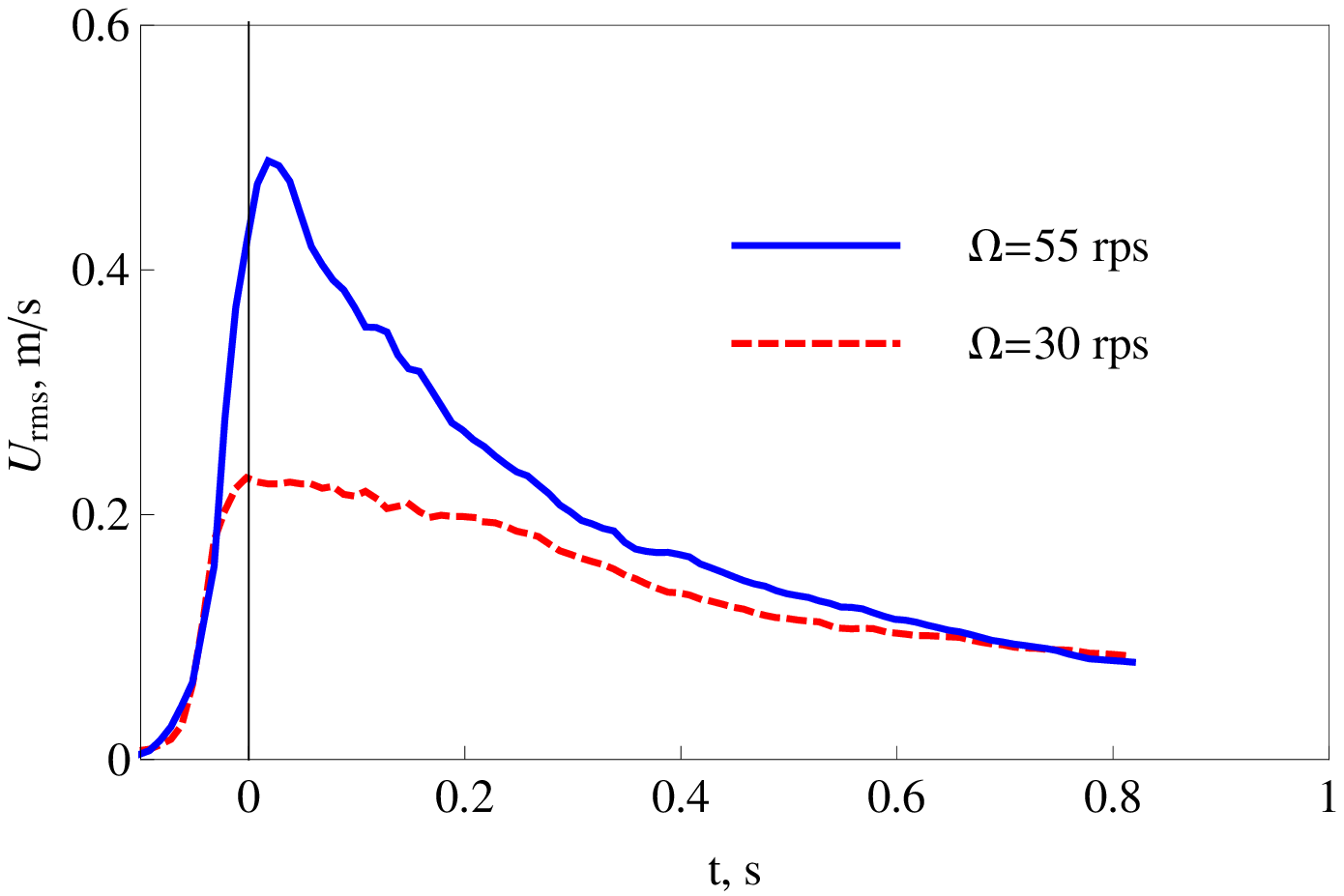}
\includegraphics[width=0.49\textwidth]{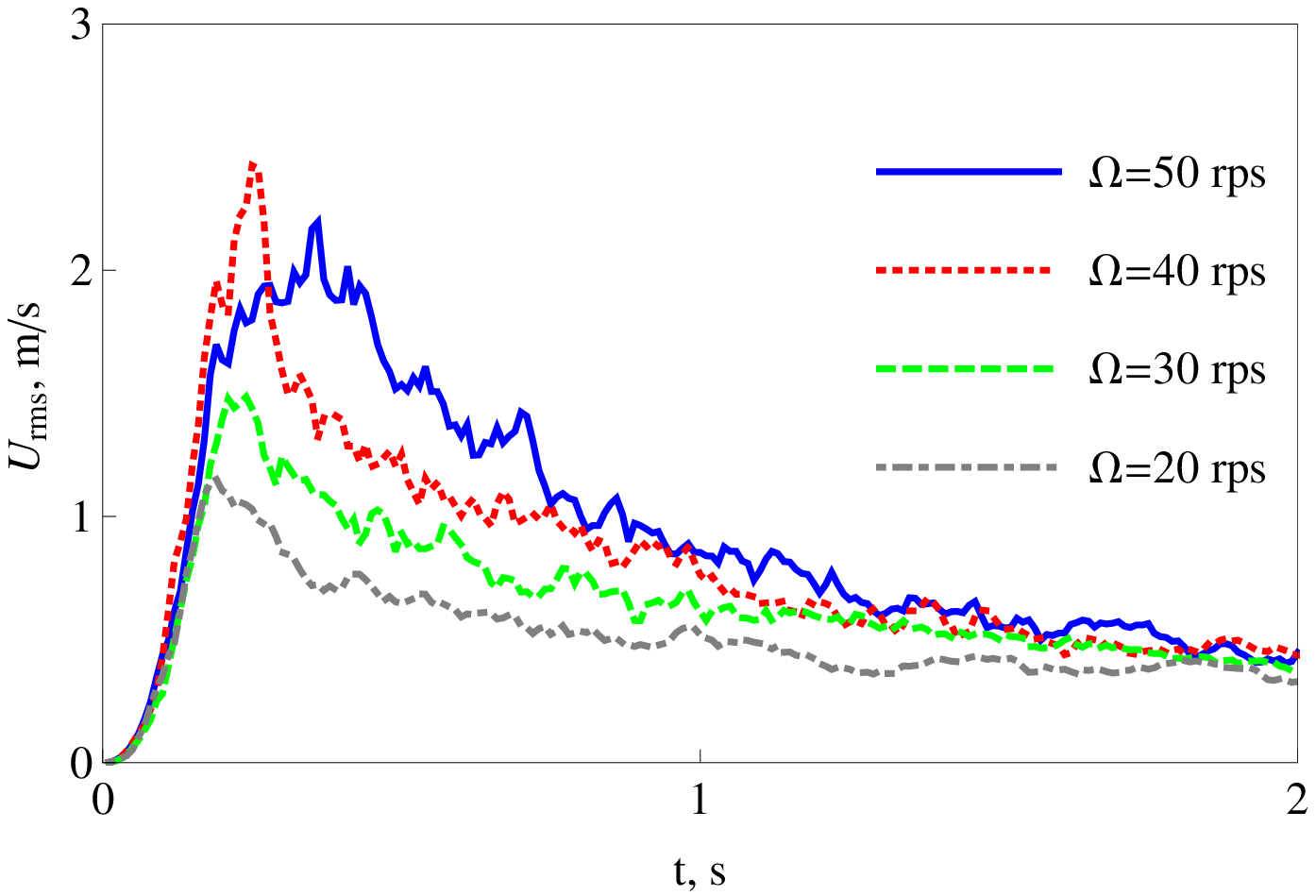}
\caption{Evolution of $U_{ rms}$ in shell-model simulations  for gallium (upper panel) and sodium (lower panel) experiments.  }
\label{fig:Uevol}
\end{center}
\end{figure}
\begin{figure}
\begin{center}
\includegraphics[width=0.49\textwidth]{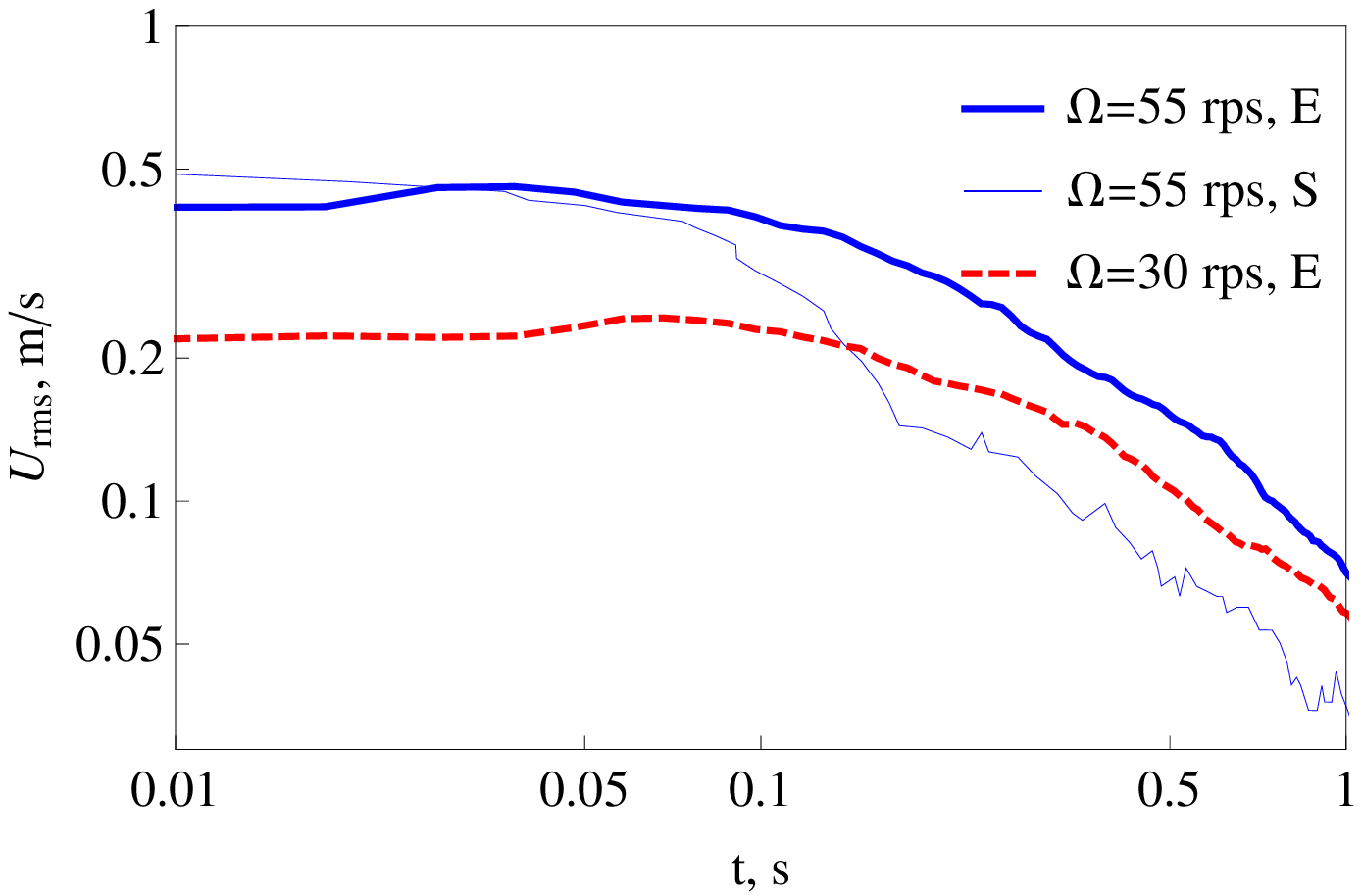}
\includegraphics[width=0.49\textwidth]{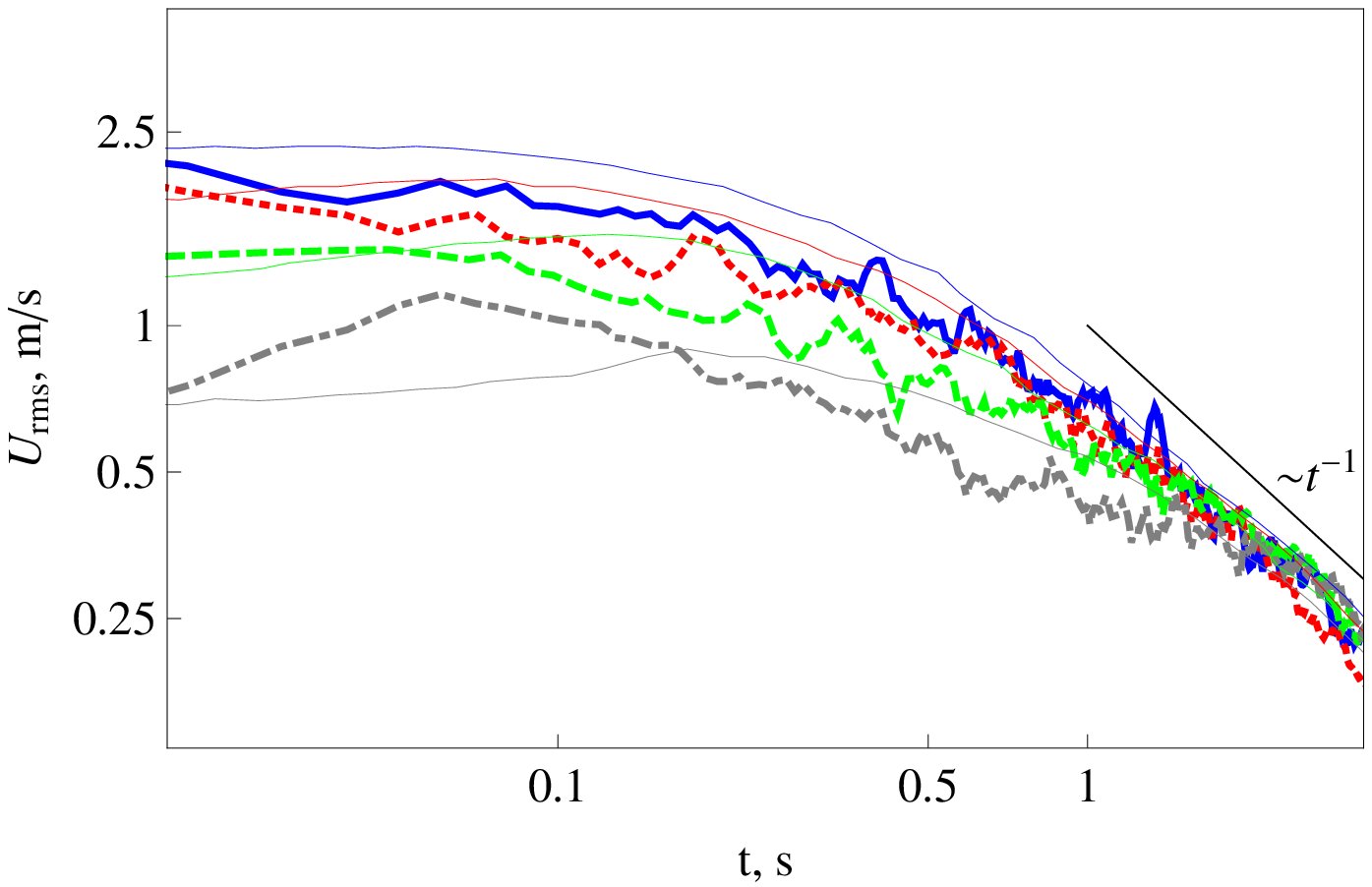}
\caption{Same as in Fig.~\ref{fig:Uevol}, but in log-log scale (thick lines) shown together with experimental data for $V_{rms}^{pol}$ (thin lines). The reference time for each regime is the end of braking. }
\label{fig:UevolLog}
\end{center}
\end{figure}

To test the capability of our model to reproduce the evolution of small-scale turbulence in spin-down flows and to estimate the empirical parameter $\kappa$ in Eq.(\ref{kappa}), we simulated Eqs.(\ref{eq}), using Eqs.(\ref{Vv})-(\ref{kappa}) and the experimental data for $V_{tor}(t)$. We examined different values of $\kappa$ and found that $\kappa=2$ provides satisfactory reproducibility of gallium results, whereas $\kappa=12$ provides the best fit for sodium experiments.

Fig.\ref{fig:Uevol} shows the evolution of rms value of velocity fluctuations in time, and the same data are plotted in Fig.~\ref{fig:UevolLog} using a log-log  scale. Hereafter, we follow the experimental papers \cite{Noskov2009,Noskov2012}, choosing as reference time  $t=0$ the start of braking for linear plots and the end of braking  for log-log plots. Simulations have been performed for the whole set of channel rotation velocity $\Omega$ presented in the corresponding experimental papers. The experimental data were averaged over 10-20 realizations of the spin-down flow with same initial conditions. Numerical simulations were done for each experimental regime as 20 realizations with same initial conditions with a weak random noise, and all results were averaged over these 20 realizations.

Comparing the numerical and experimental results, we have paid attention to the maximal energy of velocity fluctuations and the energy decay. To analyze the decay, we use the log-log plots  (Fig.~\ref{fig:UevolLog} ) and revealed  a power law like $U_{rms} \sim t^{-1}$ (i.e. the energy decays as $E(t)\sim t^{-2}$), typical for the turbulent decay.

Next, being convinced of a realistic reproduction of kinetic energy evolution, we have investigated  the temporal evolution and spectral properties of the small-scale magnetic field generated by turbulent dynamo. To simplify the comparison between velocity and magnetic field fluctuations, we have measured the magnetic field induction in terms of velocity, i.e. we plot  $B/\sqrt{\rho\mu\mu_0}$, and  magnetic field is measured in m/s.

\begin{figure}
\begin{center}
\includegraphics[width=0.49\textwidth]{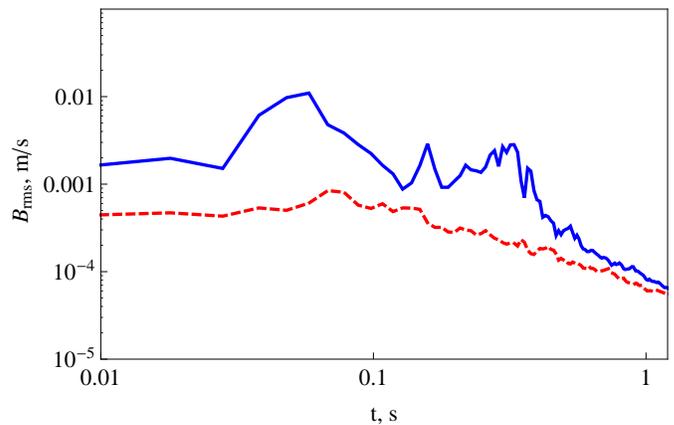}
\caption{Magnetic field evolution in the gallium spin-down flow: rms of magnetic field fluctuations  $B_{rms}$  versus time are shown in log-log scale.}
\label{fig:BevolutionGa}
\end{center}
\end{figure}
\begin{figure}
\begin{center}
\includegraphics[width=0.49\textwidth]{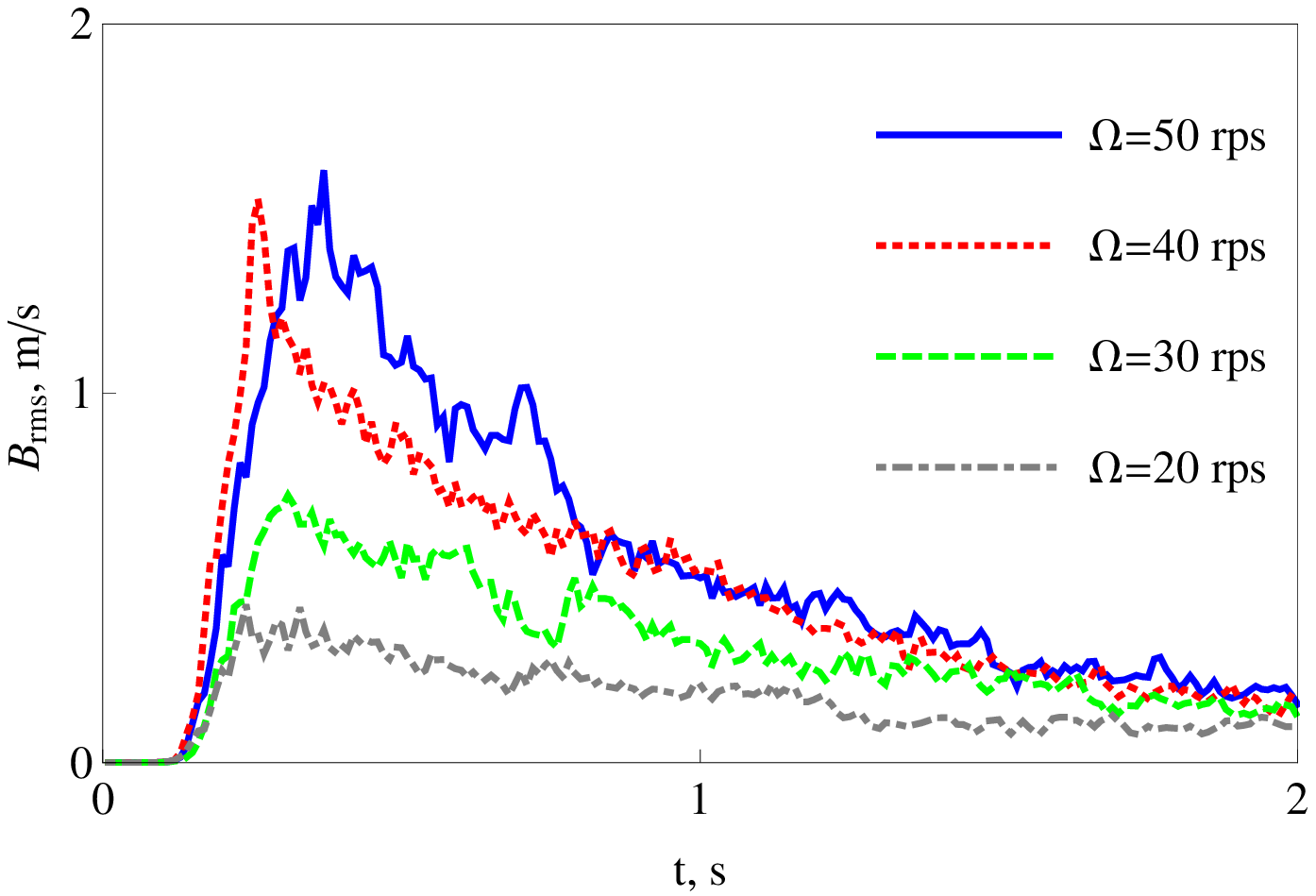}
\includegraphics[width=0.49\textwidth]{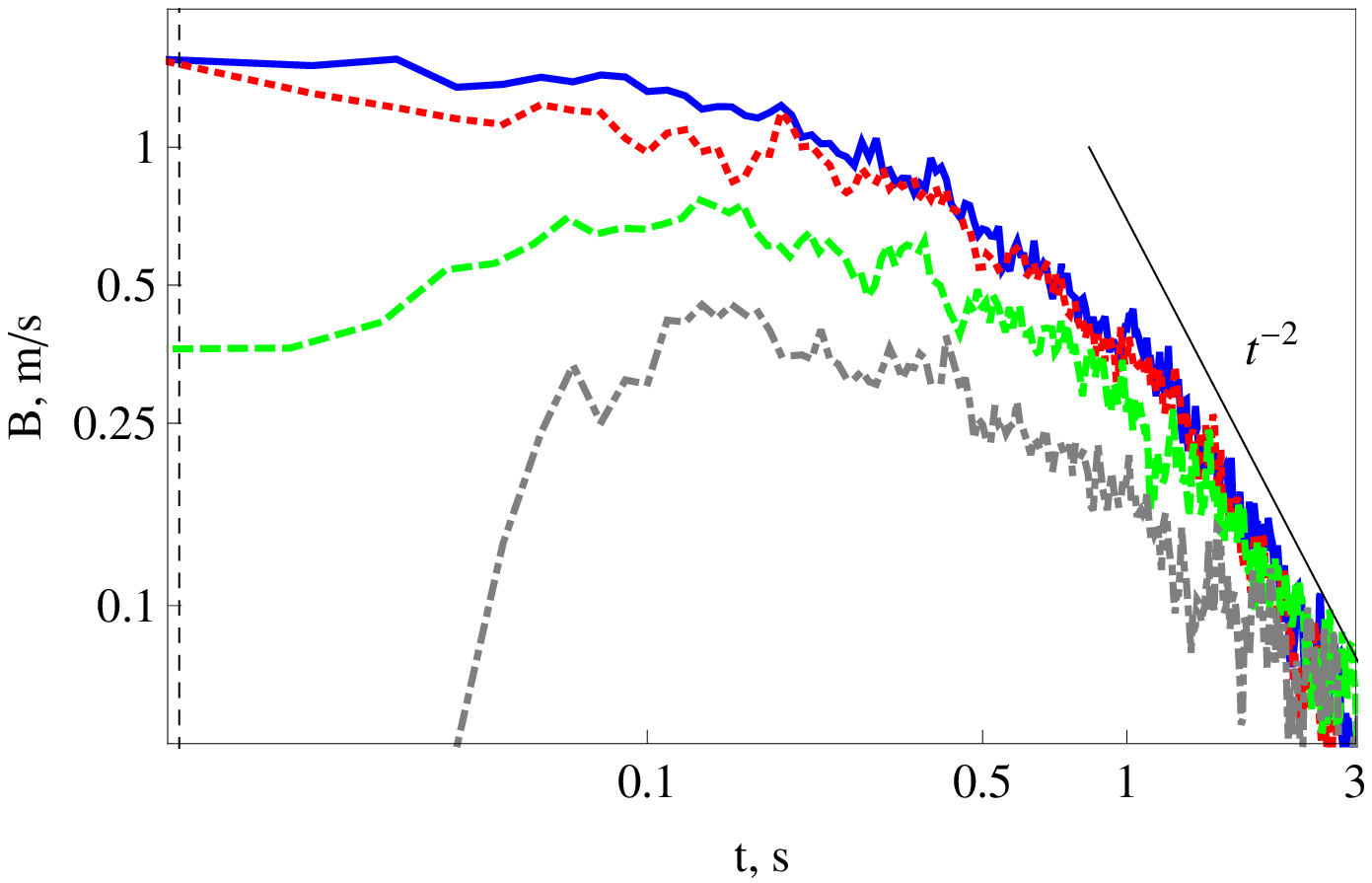}
\caption{Magnetic field evolution in the sodium spin-down flow:   $B_{rms}$  versus time in log-log scale. }
\label{fig:BevolutionNa}
\end{center}
\end{figure}

Evolution of the turbulent magnetic field $B$ in gallium spin-down flow is illustrated in Fig.\ref{fig:BevolutionGa}, where we show the rms value of magnetic field fluctuations defined as $B_{rms}=\sqrt{\sum{|B_n|^2}}$. Our simulations confirm the conclusion that only at the highest channel rotation velocity the induction effects  become observable in the turbulent gallium flow. At $\Omega=30$ rps the intensity of fluctuations is completely defined by the linear term $B_E k_n U_n$ and monotonically decays together with the flow. At $\Omega=55$ rps the maximal value of  $B_{rms}$ exceeds the level of initial (linear) fluctuations by an order of magnitude, reaching about $10^{-7}$ m/s, being about $10^{-6}$ of $U_{rms}$ only. This peak of small-scale magnetic field fluctuations is observed at the relatively late stage of the flow decay ($t \approx 0.1-0.2$s), when the kinetic energy of turbulence is  about half its maximum value. Thus, the small-scale dynamo is relatively slow and develops  in the decaying flow only.

In the sodium experiments the turbulent magnetic Reynolds number, defined through the $U_{rms}$, overcomes unity \cite{Noskov2012},  and the behavior of the magnetic field changes. The results of simulation of magnetic field evolution in the sodium spin-down flow are shown in Fig.\ref{fig:BevolutionNa}. Under slow channel rotation ($\Omega=20$ rps),  a more intense magnetic field appears also  at  the stage of flow decay ($t\approx 1$s), while at $\Omega\geq 40$~rps the magnetic field completely develops during the channel braking and reaches the maximal intensity together with the velocity fluctuation field. The maximal value of $B_{rms}$ observed in simulation for  the initial rotation velocity $\Omega= 50$ rps is close to $0.8$ m/s, which corresponds to about $U_{rms}/3$. At a late stage, the decay of magnetic field fluctuations is similar to the decay of the velocity field, tending to  ``- 2'' law.

\begin{figure}
\begin{center}
\includegraphics[width=0.47\textwidth]{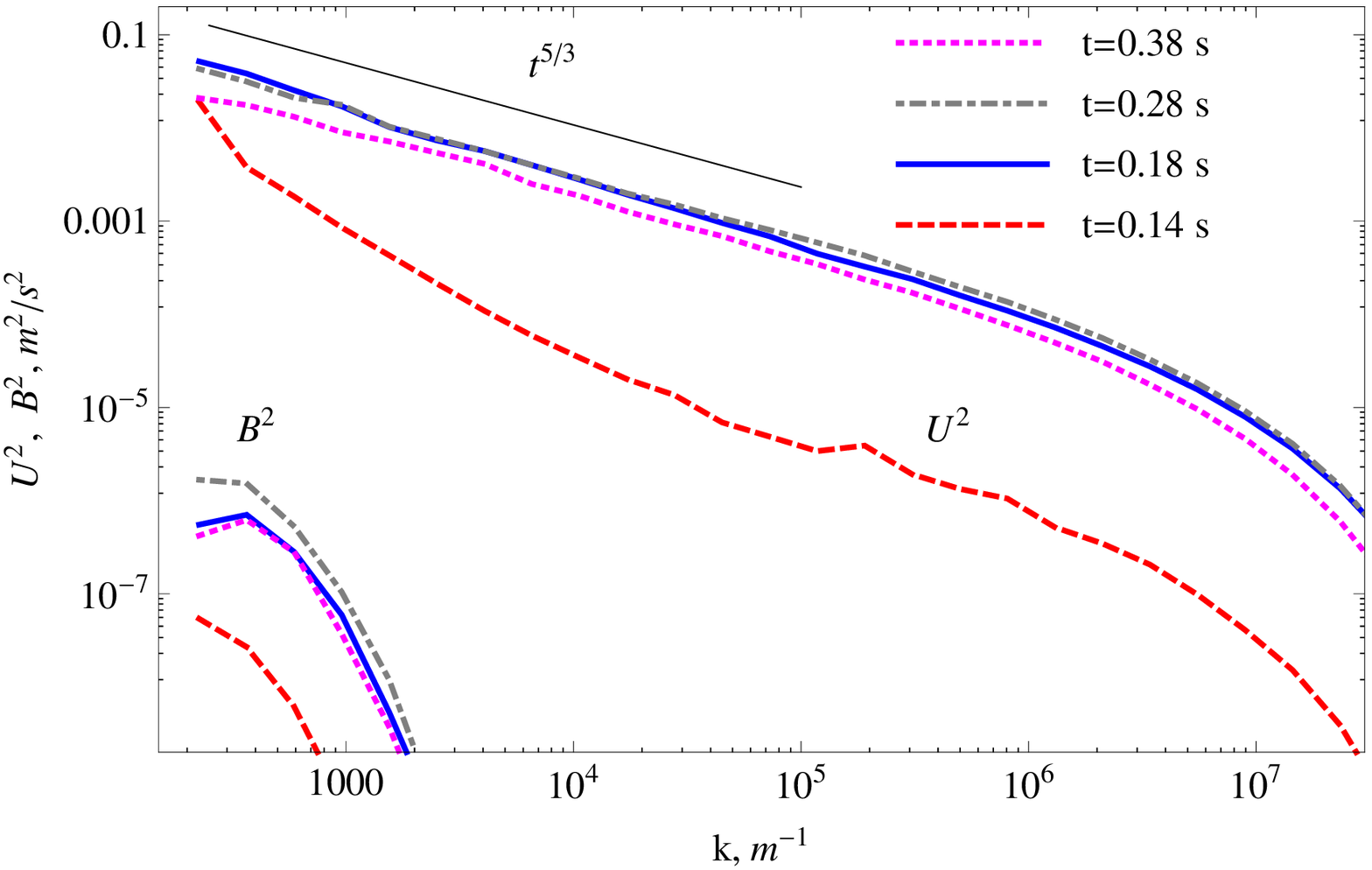}
\includegraphics[width=0.47\textwidth]{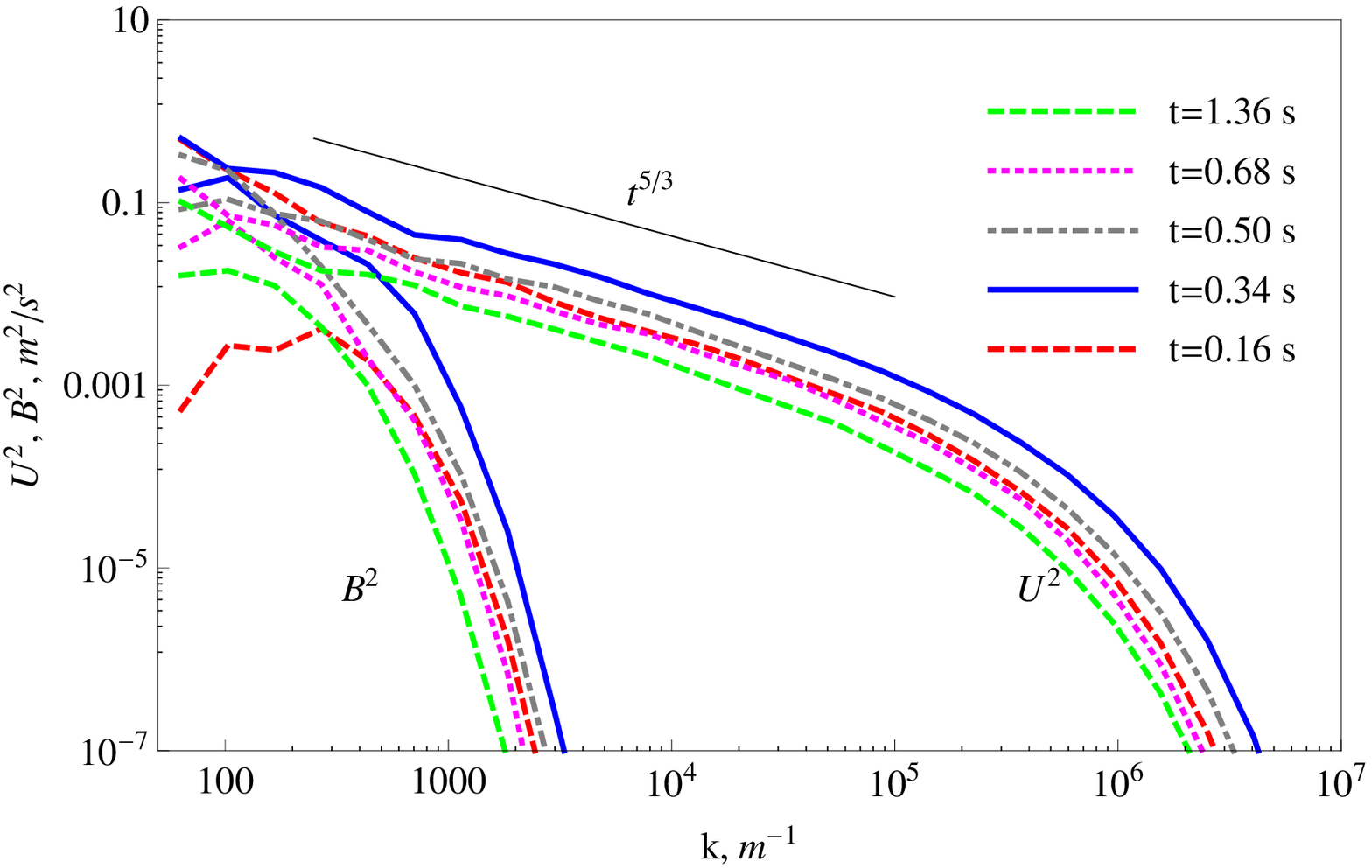}
\caption{Spectra evolution in gallium (upper panel) and sodium (lower panel) spin-down flows for initial rotation velocity $\Omega=50$  rps.
 }
\label{fig:Spec}
\end{center}
\end{figure}

Shell model simulations enable us to follow the development of spectral power density (of both velocity and magnetic fields) during the evolution of the spin-down flow. The  distribution of $U_n^2$ and $B_n^2$ at different stages of evolution of gallium and sodium flows are plotted in Fig.\ref{fig:Spec}. We show spectra for $\Omega=50$ rps during the braking (at $t=0.14$ s for gallium and $t=0.16$ s for sodium flow), at the end of braking ($t=0.18$ s and $t=0.34$ s, respectively), and at several time moments during the decay ($t=0.28, 0.38$ s for gallium and $t=0.50, 0.68, 1.36$ s for sodium). In the sodium spin-down flow, the spectrum of kinetic energy develops very fast, demonstrating that a pronounced inertial range exists already in the middle of braking and includes more than three decades of scales. In the gallium flow, the evolution of the inertial range is slower,  the developed inertial range appears at the very end of  the braking. Note that the large scale eddy turnover time, estimated as $\tau=r/U_{rms}$, at the end of the braking is similar for both flows ( $\tau\approx 0.03$ s for $\Omega=50$ rps).The higher velocity of the sodium flow is compensated by the smaller radius of the gallium channel. 

The small-scale dynamo starts in the smallest available scales which  are about $k\approx 200-500$ in both cases. In the gallium flow, the fluctuating magnetic field remains localized nearby these scales -- the $B$ spectrum looks like a narrow peak. In the sodium flow, the range of scales of magnetic fluctuations becomes more extended, and an essential part of kinetic energy is transferred to magnetic energy. This mhd-sink of kinetic energy in relatively large scales leads to shortening the inertial range. The Reynolds number in the sodium flow is higher, but the inertial range is shorter than in the gallium flow. Note that in the sodium flow under the fastest rotation ($\Omega=50$ rps) the maximal value of $U_{rms}$ is lower than at slower rotations ($\Omega=40$ rps) due to the fast development of turbulent dynamo, which consumes a visible part of kinetic energy.

\subsection{Hypothetical cases}

In this subsection, we go beyond the feasibility of the laboratory experiment, trying to find the answers to the following two questions: what would happen if we were able to rotate the toroidal channel with sodium  as fast as we want and what would happen if we were able to increase the sodium electroconductivity (magnetic Prandtl number) as high as we want?

\begin{figure}
\begin{center}
\includegraphics[width=0.49\textwidth]{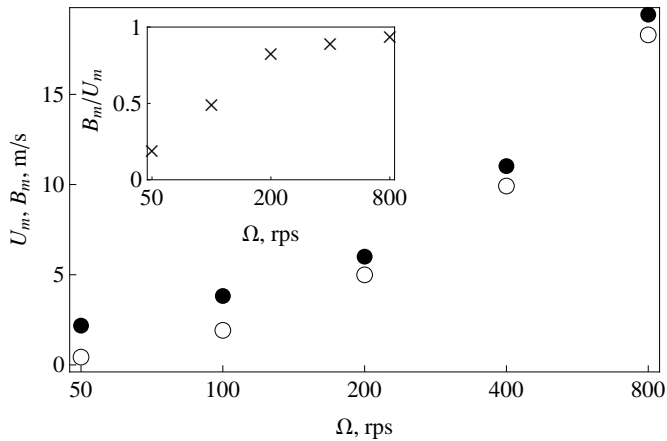}
\caption{Maximum values of $U_{rms}$ (closed circles) and $B_{rms}$ (open circles) versus $\Omega$ beyond the feasibility of laboratory experiment. $\Pm=1.15\cdot 10^{-5}$. The ratio $B_m/U_m$ is shown in the inset.
 }
\label{fig:Om}
\end{center}
\end{figure}
Concerning the first question, it is interesting to understand how far laboratory experiments are from the limit of equipartition of kinetic and magnetic energy. The results of simulations, done for sodium setup for $\Omega>50$~rps, are shown in Fig.~\ref{fig:Om}.  Note that we have no experimental data  of $V_{tor}$ for  these regimes, and threfore we use Eqs.~(\ref{ac}) -- (\ref{Vv}) to get $V_{tor}(t)$ and the corresponding $v_*(t)$. Beginning with the known regime $\Omega=50$~rps, we doubled the rotation velocity in each next numerical experiment up to $\Omega=800$~rps, increasing the attainable Reynolds number ($U_{rms}$) by a factor of ten. The generated magnetic field grows much faster. The maximal $B_{rms}$ at $\Omega=800$~rps is about 40 times larger than at $\Omega=50$~rps. The inset in Fig.~\ref{fig:Om} demonstrates that the ratio $B_{rms}/U_{rms}$  asymptotically tends to unity.

\begin{figure}
\begin{center}
\includegraphics[width=0.49\textwidth]{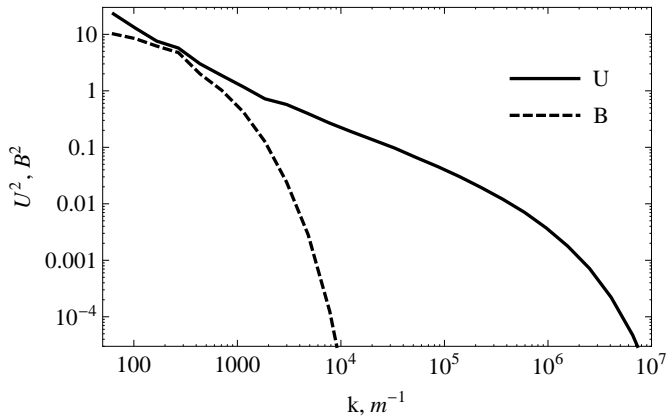}
\caption{Spectra of kinetic and magnetic energy at the end of braking  for $\Omega=100$~rps. }
\label{fig:Equip}
\end{center}
\end{figure}

The spectra shown in Fig.~\ref{fig:Equip} indicate that at $\Omega =100$~rps  the spectral density of magnetic energy approaches the spectral density of kinetic energy in a short range of scales ($200<k<400$). The low magnetic Prandtl number hampers the development of equipartition in a more extended range of scales.

\begin{figure}
\begin{center}
\includegraphics[width=0.49\textwidth]{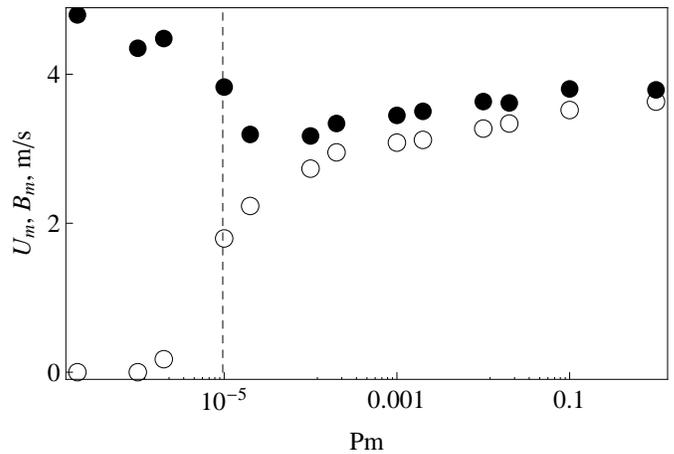}
\caption{Maximal values of $U_{rms}$ (closed circles) and $B_{rms}$ (open squares) versus the magnetic Prandtl number $\Pm$ for $\Omega=50$~rps. The sodium $Pm$ is shown by a vertical dashed line. }
\label{fig:Pm}
\end{center}
\end{figure}
Last, we have checked 
%Let us now consider 
how the efficiency of small-scale dynamo in the spin-down flow depends on the magnetic Prandtl number. The simulations were done for all real parameters of the sodium experimental setup, excepting the magnetic diffusivity, which was varied from $\eta = 10^{-6}$ to $\eta=10$ ($10^{-7}<\Pm<1$). The initial rotation velocity was fixed, $\Omega=50$~rps. Fig.~\ref{fig:Pm} shows the maximum values of $U_{rms}$ and $B_{rms}$ (denoted as $U_m$ and $B_m$) obtained in simulations with different magnetic Prandtl numbers. We see that at $\Pm\lesssim 2\cdot10^{-5}$ the dynamo is weak and no notable magnetic energy is produced. With growing $\Pm$, the dynamo becomes more and more efficient. Note that the fast growth of the magnetic field with $\Pm$ is observed in the range $2\cdot 10^{-6} \lesssim \Pm \lesssim 10^{-4}$. In this range, $U_m$ decreases, demonstrating that more and more kinetic energy is transferred into magnetic energy. Further increase of the magnetic Prandtl number leads to a slow simultaneous increase of both $U_m$ and $B_m$, provided by a reduction of Joule dissipation. The vertical dashed line in Fig.~\ref{fig:Pm} indicates the $\Pm$ of liquid sodium nearby the melting point.

\section{Discussion and Conclusions}
\label{sec:disc}

Dynamo in non-stationary flows is a particular problem, which is of special interest in the context of different astrophysical objects.
Intense pulse flows provide the possibility for reaching high magnetic Reynolds numbers in relatively small experimental setups using moderate mass of liquid metal. The spin-down flow in toroidal channels is the first flow configuration used to study dynamo effects in a non-stationary statement of the problem. In this paper,  we have made an attempt to estimate the effect of small-scale dynamo in spin-down flows of liquid metals realized in laboratory experiments \cite{Denisov2008,Noskov2009,Frick2010,Noskov2012}. Our simulations confirmed the conclusion that in gallium experiments \cite{Denisov2008,Noskov2009} the dynamo effects are weak -- a slight burst of small-scale magnetic energy arose only at the highest available rotation velocity of the channel. In the case of sodium flows,  the induction effects are quite strong \cite{Frick2010,Noskov2012}. Our simulations have shown that an essential part of kinetic energy of sodium spin-down flows is converted into magnetic energy and dissipates because of Joule heat losses.

 Specific character of the turbulent MHD system under consideration is provided by the very short and strong power supply. Then the scenario of evolution depends on the ratio between the characteristic times -- time of braking, time of inertial range formation,and time of small-scale dynamo development. If the dynamo develops rapidly consuming a visible part of kinetic energy, the kinetic energy in an inertial range is reduced in spite of lack of any magnetic field in this range of scales. An increase in the  Reynolds number does not necessarily result in the extension of the apparent inertial range. In the gallium experiment, the Reynolds number was about three time less than in the sodium one, but the inertial range was more extended.

At the highest available rotation velocity, $\Omega=50$~rps, the rms value  of magnetic field fluctuations reached about 25\% of velocity fluctuations. The range of scales of generated magnetic field remains narrow due to the small magnetic Prandtl number, but the peak of the magnetic spectral power density $E_B(k)$ is about to reach the level of kinetic $E_U(k)$. In our simulations we went far beyond the existing laboratory facilities and examined the spin-down flows under rotation velocity $\Omega \gg 50$~rps. We have found that $\Omega\approx 100$~rps is enough to reach the equipartition of $E_B(k)$ and $E_U(k)$ at the lowest wave numbers (largest scales), while under $\Omega \gtrsim 200$~rps the intensities of magnetic field and velocity field fluctuations become comparable. Thereby we have shown that the existing experimental setups are ready to reach regimes, in which the equipartition of magnetic and kinetic energy can be established at least in a short range of scales.
Analysis of the simulations results has revealed that the amplitude of the fluctuating magnetic field inside the turbulent bulk can reach  100~G in the actual sodium experiments. Today,  measurements of the local magnetic field inside the toroidal channel are not available. Our simulations show that an attempt to perform them may pose real challenges.

Finally, we explored the influence of the magnetic Prandtl number on the efficiency of small-scale dynamo in spin-down flows. The role of the magnetic Prandtl number has been largely discussed in context of different dynamo problem.  The extended DNS \cite{Schekochihin2004,Iskakov2007,Schober2012}  and the shell-model studies \cite{Stepanov2006a,Frick2006} of stationary MHD-turbulence revealed the critical level of $\Rm\approx100$, above which the real non-linear small-scale dynamo is sustainable. However, some kinematic dynamos remain possible. DNS at small $\Pm$   indicate that the magnetic energy growth takes place mainly due to the energy transfer from large-scale velocity field to large-scale magnetic field and that a weak forward magnetic energy flux exists \cite{Kumar2015}. In the experimental spin-down flows under discussion, the small-scale dynamo remains in the quasi-kinematic regime, where the main magnetic energy is dissipated into the scale, wherein it is converted from the kinetic energy. The real small-scale dynamo can start to operate at $\Pm>10^{-4}$, and an inertial range appears in the spectrum of magnetic energy. Then, the energy dissipation is postponed to later times and smaller scales, and the peak of turbulent energy (both, kinetic and magnetic) slightly increases with $\Pm$ (see Fig.~\ref{fig:Pm}).

The analysis of a contribution of viscous and Joule dissipation to energy dissipation rate in stationary forced MHD turbulence has shown that the ratio of magnetic-to-kinetic energy dissipation has a strong maximum in the dependence on the magnetic Prandtl number\cite{Plunian2010}. The localization of this maximum depends on the Reynolds number, but for $\Ru\sim 10^5--10^6$ (which is of the order of $\Ru$ in our spin-down flows) the maximum is found at $\Pm \sim 10^{-4}$, and the magnetic dissipation is one order of magnitude higher than the viscous one. It should be noted that the $\Pm$ of liquid sodium, which is recognized to be the best metal for laboratory dynamo experiments, belongs to the range of $\Pm$ with the steepest dependence of the generated magnetic energy on $\Pm$ (see Fig.~\ref{fig:Pm}). This means that in the experiments it is worth to fight for even small increase of $\Pm$ (which decreases with temperature).

Shell models are known as an effective low-resource-consuming technique for studying {\it homogeneous, isotropic, stationary} turbulence. The problem under discussion is neither homogeneous nor isotropic and, particularly, nonstationary. Nevertheless, we have made an attempt to use a shell model for estimation of small-scale dynamo effects in spin-down flows, and we believe that our estimations are reasonable. Shell models fail to provide any information concerning the spatial distribution of the characteristics of turbulent flows, and they are inefficient for  absolute estimations of turbulent characteristics.  On the contrary, they are efficient for studying relative characteristics, such as spectral distribution of energy components, helicities, enstrophy, parts of spectral fluxes, scale-by-scale ratio of different quantities, etc. \cite{Plunian2013}. This allows us to hope that after calibration against the experimental data for turbulent kinetic energy, the shell model provides an adequate estimation of the evolution of small-scale magnetic energy during the development and decay of a spin-down flow.

\section*{Acknowledgment}

The work was supported by the Russian foundation for basic research and Perm regional government under the project RFBR-17-41-590059.

\bibliography{references}
\end{document}